\documentclass[journal]{IEEEtran}
\ifCLASSINFOpdf
    \usepackage[pdftex]{graphicx}
\else
\fi
\usepackage{amsmath}
\begin{document}
%
\title{Hardware Implementation of a Fast Algorithm for the Reconstruction of Muon Tracks in ATLAS Muon Drift-Tube Chambers for the First-Level Muon Trigger at the HL-LHC}
%
%
%

\author{\IEEEauthorblockN{S.~Abovyan},
\IEEEauthorblockN{V.~Danielyan},
\IEEEauthorblockN{M.~Fras},
\IEEEauthorblockN{Ph.~Gadow},
\IEEEauthorblockN{O.~Kortner},
\IEEEauthorblockN{S.~Kortner},
\IEEEauthorblockN{H.~Kroha},
\IEEEauthorblockN{F.~M\"uller},
\IEEEauthorblockN{S.~Nowak},
\IEEEauthorblockN{R.~Richter},
\IEEEauthorblockN{K.~Schmidt-Sommerfeld}\\
\IEEEauthorblockA{Max-Planck-Institut f\"ur Physik, F\"ohringer Ring 6, 80805
M\"unchen, Germany}}

%
%

\markboth{Journal of \LaTeX\ Class Files,~Vol.~14, No.~8, August~2015}%
{Shell \MakeLowercase{\textit{et al.}}: Bare Demo of IEEEtran.cls for IEEE Journals}
%



\maketitle

\begin{abstract}
The High-Luminosity LHC will provide the unique opportunity to explore the nature of physics beyond the Standard Model of strong and electroweak interactions. Highly selective first level triggers are essential for the physics programme of the ATLAS experiment at the HL-LHC where the instantaneous luminosity will exceed the LHC Run 1 instantaneous luminosity by almost an order of magnitude. The ATLAS first level muon trigger rate is dominated by low momentum muons, selected due to the moderate momentum resolution of the resistive plate and thin gap trigger chambers. This limitation can be overcome by including the data of the precision muon drift tube (MDT) chambers in the first level trigger decision. This requires the fast continuous transfer of the MDT hits to the off-detector trigger logic and a fast track reconstruction algorithm performed in the trigger logic.

In order to demonstrate the feasibility of reconstructing tracks in MDT chambers within the short available first-level trigger latency of about 3~$\mu$s we implemented a seeded Hough transform on the ARM Cortex A9 microprocessor of a Xilinx Zynq FPGA and studied its performance with test-beam data recorded in CERN's Gamma Irradiation Facility. We could show that by using the ARM processor's Neon Single Instruction Multiple Data Engine to carry out 4 floating point operations in parallel the challenging latency requirement can be matched.

\end{abstract}


%
\IEEEpeerreviewmaketitle

\section{Introduction}
%
%
%
%
\IEEEPARstart{T}{he} trigger of the ATLAS experiment at the Large Hadron Collider uses a three-level trigger system. The first-level (L0) trigger for muons with high transverse momentum $p_\mathrm{T}$ is based on chambers with excellent time resolution (better than 20 ns), able to identify muons coming from a particular beam crossing.

The trigger chambers also provide a fast muon $p_\mathrm{T}$ measurement, however with limited accuracy due to their moderate spatial resolution along the deflecting direction of the magnetic field. The limited momentum resolution weakens the selectivity of the L0 trigger for high $p_\mathrm{T}$ muons above a predefined threshold, like 20~GeV, accepting muons below the threshold.

The higher luminosity foreseen for the phase II of the LHC, the so-called ``High-Luminosity LHC'', puts stringent limits on the L0 trigger rates. A way to control these rates would be to improve the spatial resolution of the triggering system resulting in a drastically sharpened turn-on curve of the L0
trigger with respect to $p_\mathrm{T}$. This is possible without the installation of new trigger chambers with higher spatial resolution by complementing the position measurements of the existing trigger chambers with the much more precise position measurements of the monitored drift-tube (MDT) chambers which are installed in the ATLAS detector to provide an accurate muon momentum measurement.

In this concept the trigger chambers will be used to define regions of interest (RoI) inside which high $p_\mathrm{T}$ muon candidates have been identified. MDT hits in the RoI(s) are passed to the trigger logic, where they are used for an accurate estimate of the track momentum, leading to an efficient suppression of sub-threshold muon triggers. 


%
%

\section{The ATLAS muon spectrometer at the HL-LHC}
At the LHC the ATLAS muon spectrometer uses three layers of muon chambers operated in a magnetic field created by an air-core toroid system to trigger on  muons with high transverse momenta $p_\mathrm{T}$ up to a pseudorapidity $|\eta|=2.4$ and to measure $p_\mathrm{T}$ with 4\% resolution in a wide momentum range and 10\% at $p_\mathrm{T}=1$~TeV up to $|\eta|=2.7$ \cite{AtlasDetectorPaper}. Resistive Plate Chambers (RPC) in the barrel region and Thin Gap Chambers (TGC) in the end-cap regions with excellent time resolution of a few nanoseconds for $pp$ bunch crossing identification, but moderate spatial resolution are used for the first level trigger. The high muon momentum resolution is achived by a precise measurement of muon trajectories with Cathode Strip Chambers (CSC) in the inner end-cap disk at large rapidities and Monitored Drift Tube (MDT) chambers in the rest of the spectrometer. These chambers have spatial resolutions better than 40~$\mu$m.

ATLAS uses a 3-level trigger system. A high-$p_\mathrm{T}$ muon trigger built out of a conincidence of three RPCs in the barrel toroid magnet and three TGCs behind the end-cap toroid magnet is part of the first trigger level. The muon momentum is estimated from the sizes of the deviations of the trigger chamber hits from a straight line from the $pp$ interaction point. 

The ATLAS muon spectrometer is operated in a large background of neutrons and $\gamma$ rays. In order to cope with  background counting rates of up to 15~kHz\,cm$^{-2}$ the so-called ``small wheel'' will be replaced by a new small wheel (NSW)with chambers with increased high-rate capability in the long shutdown 2 \cite{Kawamoto:1552862}. The other muon spectrometer upgrades will be carried out in long shutdown 3. In the NSW small strip TGCs (sTGCs) will be used for triggering while MicroMegas will be used for precision tracking and a refined muon momentum measurement at trigger level. The big wheel's TGCs closest to the beam pipe will be replaced with TGCs with higher spatial resolution to increase the selectivity of the 1$^{st}$ level muon trigger. New thin-gap RPCs will be added to the inner barrel layer to close acceptance gaps of the barrel muon trigger. To free space for the RPCs MDT chambers will need to be replaced by so-called ``sMDT chambers'' which are drift-tube chambers with 15~mm diameter tubes instead of 30~mm diameter tubes.  Finally the new trigger architechture will require new on- and off-chamber electronics. \cite{muon_phaseII_TDR}

\section{Data flow}
\begin{figure}[hbt]
\begin{center}
    \includegraphics[width=\linewidth]{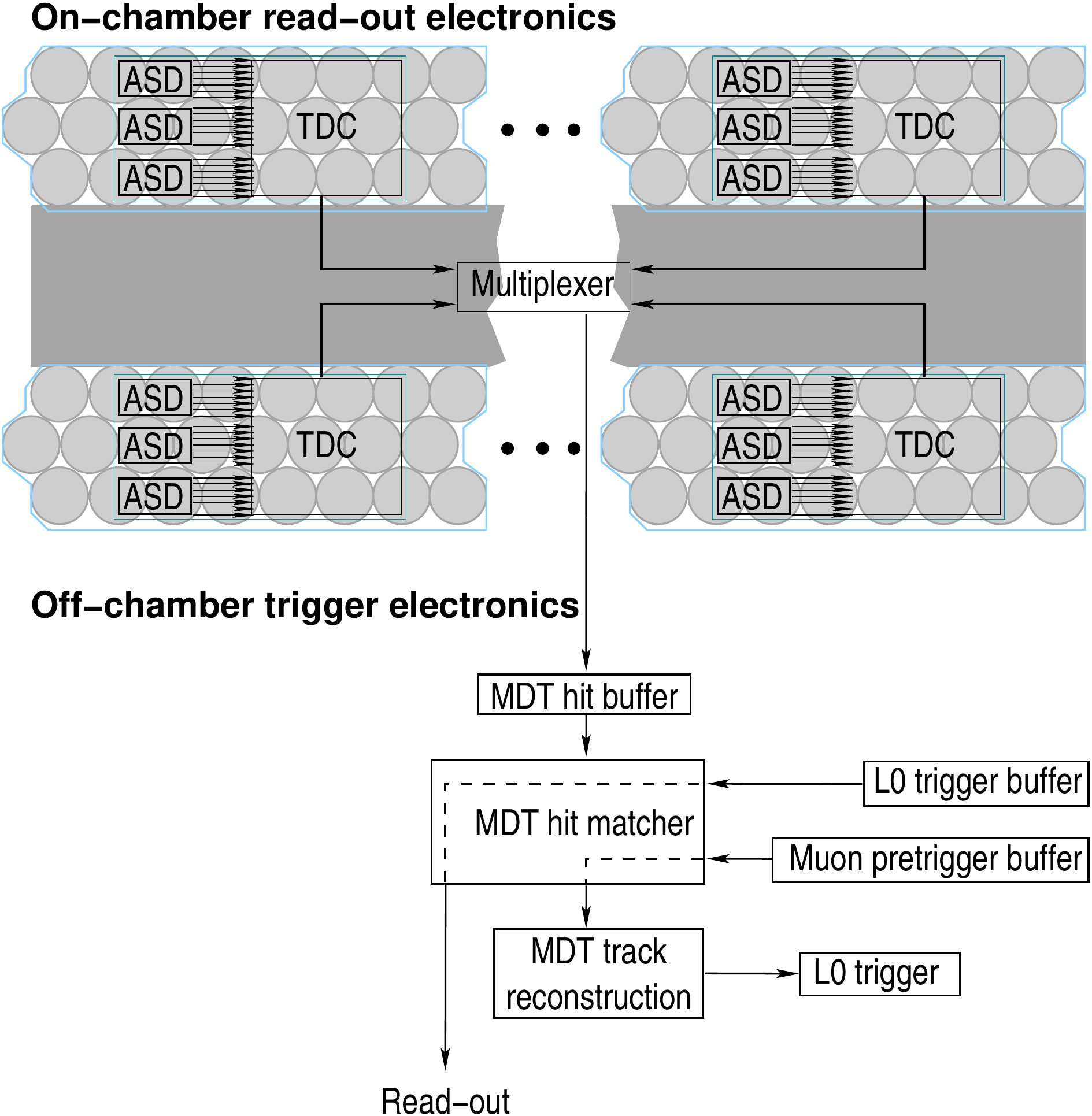}
    \caption{\label{sec:dataflow:fig01} Schematic drawing of the functional units of the MDT chamber read-out and the off-chamber MDT trigger electronics.}
\end{center}
\end{figure}

In order to maximize the time which is available for the muon momentum determination from the MDT data one needs to minimize the time for the transmission of the MDT data to the off-chamber trigger electronics. Figure~\ref{sec:dataflow:fig01} shows a diagram of the functional units which are needed to digitize the MDT hits in the on-chamber electronics and to send the MDT data off to the trigger electronics. The on-chamber electronics consists of an amplifier, shaper, and discriminator for each drift tube for hit detection. The output of the discriminator is connected to a time-to-digital converter (TDC) which measures the hit time with a clock synchronized to the LHC bunch crossing frequency with 0.78125~ns precision. The hit time and the tube number can be stored in a single 32-bit word. The 32-bit hit words are collected by an on-chamber multiplexer which transmits the 32-bit hit words to the trigger electronics over an optical fibre.

The $\gamma$-ray background in the ATLAS muon spectrometer is the main source of MDT hits and leads to a hit rate of up to 200~kHz per tube. This corresponds to a data rate of 6.4~Mbps per tube and to a maximum rate of 2.8~Gbps for the largest chambers with 432 drift tubes. One will therefore have to use gigabit transmission of the MDT data from the on-chamber multiplexer to the trigger electronics.

In the off-detector MDT trigger electronics the MDT data will have to be buffered for 10~$\mu$s which is the maximum latency of the first-level trigger. As the on-chamber TDC is operated in triggerless mode to minimize the overall latency, the MDT hits must be matched in time to L0 triggers which come at an average maximum rate of 1~MHz and are stored in another buffer in the trigger electronics. The MDT data which can be matched to a L0 trigger are sent off to the read-out system of the ATLAS detector. In order to be able to refine the momentum measurement of the trigger chamber system with the spatial precision measurements of the MDT chambers one must implement a separate processing path of the MDT data for the muon trigger candidates which are found by the trigger chamber system and will be called ``muon pretriggers'' from now on. On a muon pretrigger the MDT hits are matched in time and space to the pretrigger, the drift times of the matched hits are converted into drift radii, and the matched hits are sent to the MDT track reconstruction unit for the refinement of the muon momentum measurement as the basis of the muon trigger decision.

In the track reconstruction unit the following processing steps are carried out:
\begin{enumerate}
    \item   Straight-line track segments are reconstructed in each muon chamber
            in the region of interest of the pretrigger.
    \item   If track segments were found in each of the three muon chamber
            layers of the muon spectrometer the positions of the three segments
            are used to compute the sagitta $s$. In the barrel part of the      
            spectrometer $S$ is defined as the distance of the position of the 
            segment in the middle chamber from the straight-line 
            interconnections of the segment positions in the inner and the outer
            chamber. In the end caps $s$ is defined a the distance of the
            segment position in the small wheel from the straight line through
            the segment positions in the big wheel and the outer wheel. If track             segment have only beend found in two barrel chamber layers or only
            in the big or outer end-cap wheel the angle $\beta$
            between the segments is computed.
    \item   $s$ and $\beta$ measure the deflection of the muon trajectory in
            magnetic field of the muon spectrometer and are therefore used
            for the measurement of the muon momentum.
\end{enumerate}

\section{MDT trigger hardware}
The MDT chambers are arranged in 4 groups of 16 sectors: in the negative hemisphere of the spectrometer one group in the end cap and one group in the barrel and the same arrangement in the positive hemisphere. Each sector contains three layers of 6 MDT chambers. The expected rate of muon pretriggers with a transverse momentum threshold of 20~GeV/c is less than 3~kHz per sector corresponding to an average time between two consecutive pretriggers of 0.3~ms which is large compared to the maximum L0 latency of 10~$\mu$s. So one MDT trigger processor board per sector is a comfortable and natural design choice.

\begin{figure}[hbt]
\begin{center}
    \includegraphics[width=\linewidth]{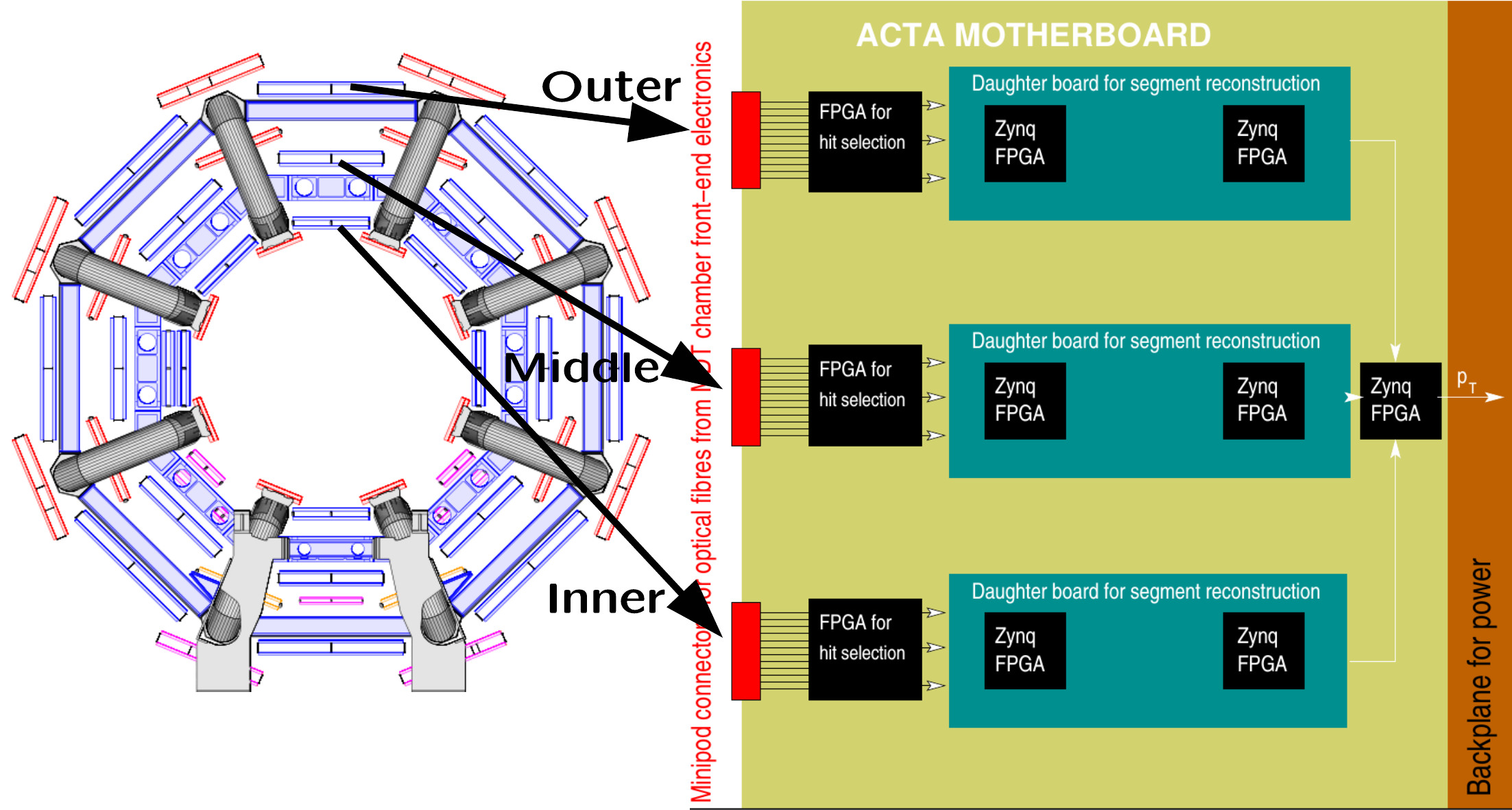}
    \caption{\label{sec:hardware:fig01} Schematic conceptual drawing of an ATCA blade for the MDT trigger with the cross section of the muon spectrometer barrel illustratign the assignment of an ATCA blade to a spectrometer sector.}
\end{center}
\end{figure}

Figure~\ref{sec:hardware:fig01} summarizes this hardware segmentatio pictorially. The MDT data of a sector can be sent via optical fibres to an ATCA blade with one FPGA per chamber layer for the MDT hit matcher. For the segment reconstruction we propose to use an FPGA with embedded microprocessor like a Xilinx Zynq FPGA which can be purchased with a dual core ARM Cortex A9 processor running at 1~GHz clock frequency. As will be explained in the following section, the FPGA core will be used for pattern recognition where parallel processing is required, the floating point arithmetics for the determination of the segment parameters will be done on the ARM processor. In our proposal one ARM core is used per segment. In order to handle up to four muon pretriggers in a sector we need two Zynq FPGAs per chamber layer, six Zynq FPGAs in total. It seems most convenient to us to put the two Zynq FPGAs for the segment reconstruction in a chamber layer on a daughter board. The parameters of the segments reconstructed in the three daughter boards have to be collected in separate small Zynq FPGA for the determination of the muon momentum. This Zynq FPGA will send the value of the muon momentum to the higher level muon trigger electronics.

\section{A fast and lightweight segment reconstruction algorithm}
In the segment reconstruction processes one must identify the MDT hits which lie on a common straight muon trajectory and determine the parameters of this trajectory. In this article we shall call the first part of the segment reconstruction pattern recognition and the segment part segment fit.

\begin{figure}[hbt]
\begin{center}
    \includegraphics[width=0.4\linewidth]{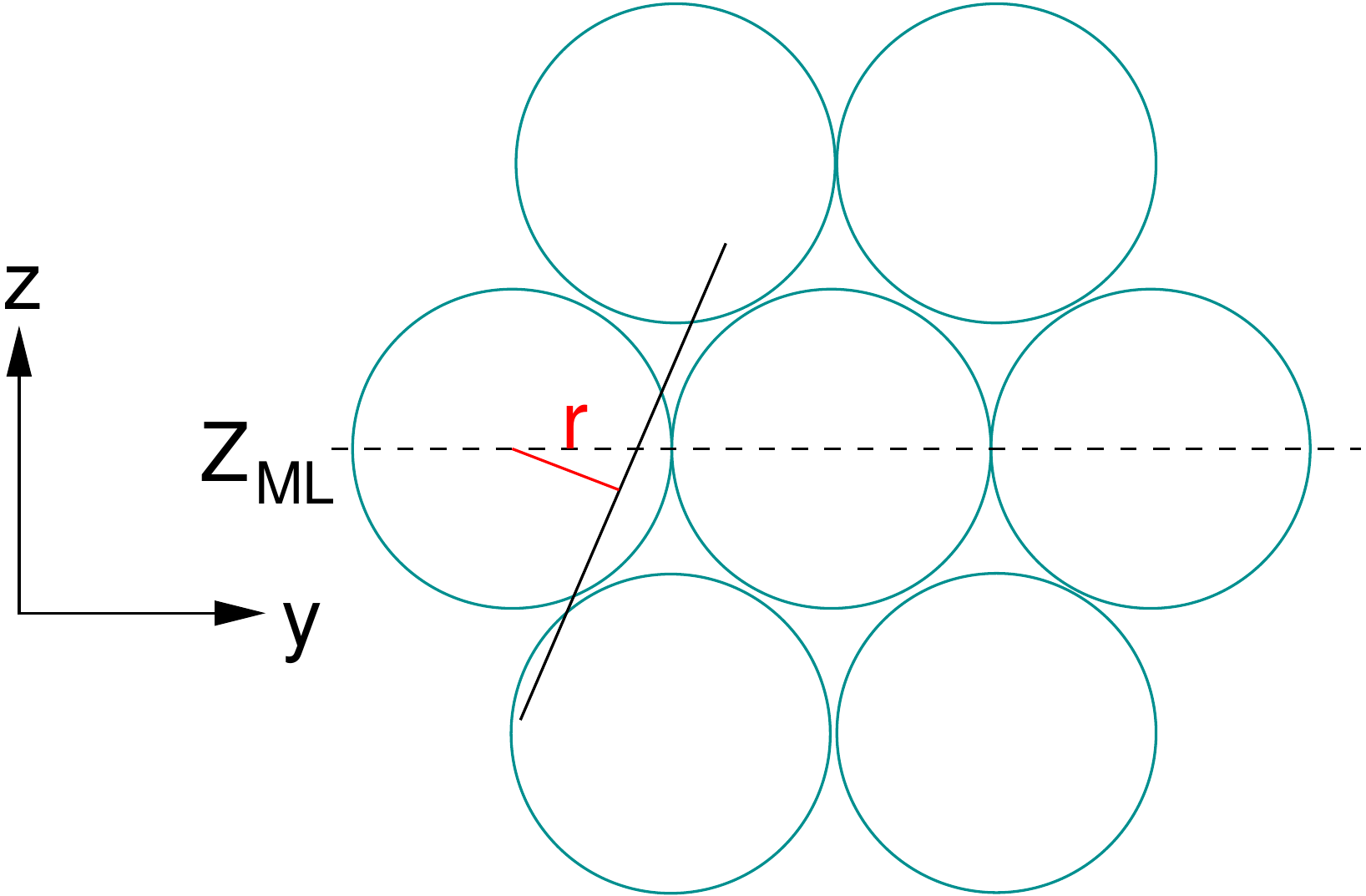}
    \caption{\label{sec:recalg:fig01} Schematic drawing illustrating the
    definition of the chamber coordinate system and the distance $r$ of straight
    line from the anode wire of a drift tube.}
\end{center}
\end{figure}

For the following discussion we are using a coordinate system in which the $z$~axis is orthogonal to the drift tube layers and the $y$~axis is parallel to the tube layers (see Figure~\ref{sec:recalg:fig01}). If $(y,z)$ denotes the position of the anode wire of a tube, the distance of a straight line
\begin{eqnarray}
    y = m\cdot z+b
    \nonumber
\end{eqnarray}
with slope $m$ and intercept $b$ from this anode wire is given by
\begin{eqnarray}
    r = \frac{|m\cdot z+b-y|}{\sqrt{1+m^2}}.
    \nonumber
\end{eqnarray}
The slope $m$ is measured by the trigger chamber system with an accuracy between 4~mrad in the end caps and about 20~mrad in the barrel part of the muon spectrometer. Using the drift radius measured by the MDT chamber and the slope $\bar{m}$ provided by the trigger chamber system one can transform this equation into an equation for the intercept $b$
\begin{eqnarray}
    b_\pm = r\cdot\sqrt{1+\bar{m}^2}-(\bar{m}\cdot z-y)
    \nonumber
\end{eqnarray}
where $b_\pm$ denotes the two solutions for the intercept. The precision of $b_\pm$ is dominated by the moderate resolution of $\bar{m}$ and depends on the choice of the origin of the coordinate system. If one choses the origin of the coordinate system in the middle of the multilayer in which the MDT hit was detected, the precision of $b_\pm$ is better than about 1~mm. A resolution of 1~mm is sufficient to idenficaty the MDT hits which belong to the same straight line and on which side of the anode wire the muon traversed the tube.

In our algorithm we compute $b_\pm$ for each hit in the region of interest in multiples of 1~mm and sort the $b_\pm$ values in increasing order. We go throught the sorted list and count how many times each value occurs to find the largest accumulation of equal $b_\pm$ values. The hits which are associated to the largest accumulation are the hits which lie on a common straight line. 

\begin{figure}[hbt]
\begin{center}
    \includegraphics[width=0.3\linewidth]{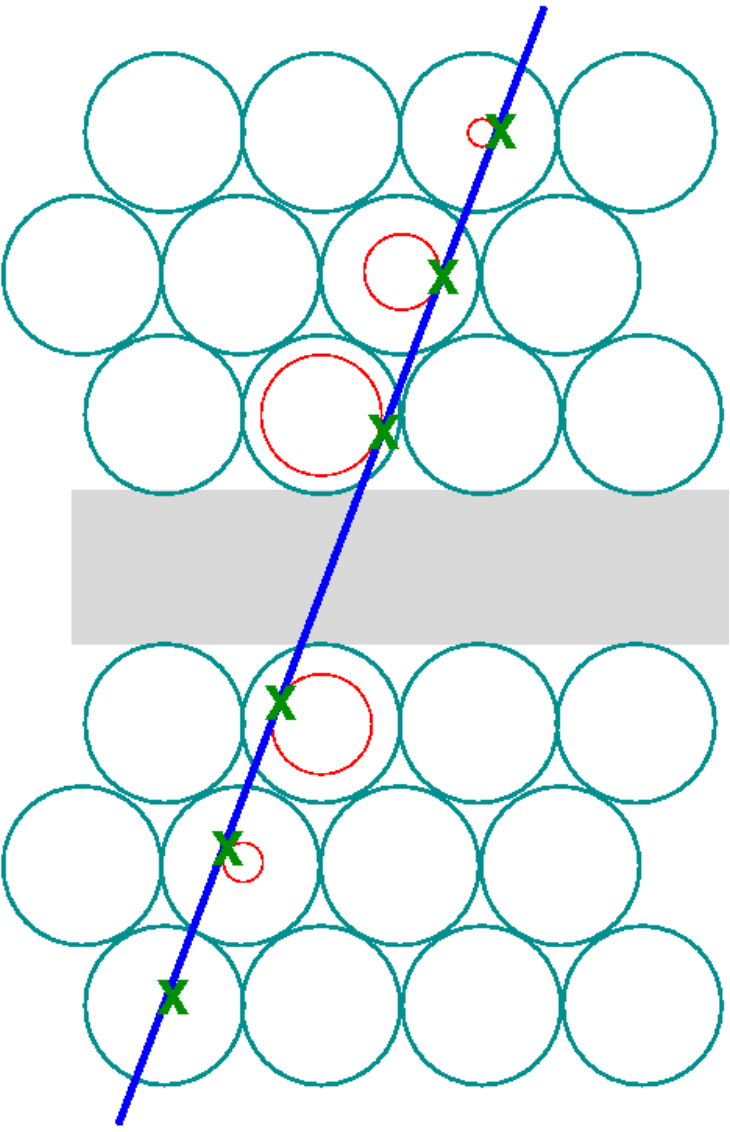}
    \caption{\label{sec:recalg:fig02} Schematic drawing illustrating the
    situation before and after the pattern recognition. Before the pattern
    recognition one has drift radii and the tubes. After the pattern 
    recognition one knows which tubes have been traversed by the straight muon
    trajectory and which space points on this straight line have been measured
    by the tubes.}
\end{center}
\end{figure}

For each of these hits we know the sign $alpha$ in front of the $r\sqrt{1+m^2}$ term. So if we replace the slope $m$ by $\bar{m}+\delta$ and linearize $\sqrt{1+(\bar{m}+\delta)^2}$ in $\delta$, we get a single equation
with two free parameters, $\delta$ and $b$ for each hit:
\begin{eqnarray}
    \alpha r\sqrt{1+\bar{m}^2}-(\bar{m}z-y)+
    \left(\frac{\alpha r\bar{m}}{\sqrt{1+\bar{m}^2}}-z\right)\delta - b = 0.
    \nonumber
\end{eqnarray}

$\delta$ and $b$ can be determined by a least-square fit to the $n$ points

$\left(\alpha_k r_k\sqrt{1+\bar{m}^2}-(\bar{m}z_k-y_k), \frac{alpha_k r_k \bar{m}}{\sqrt(1+\bar{m}^2)}-z_k\right)_{k\in\{1,\dots,n\}}$.

\section{Momentum determination}
The deflection of the muon trajectory in the magnetic field can be measured by the angle $\beta$ between two MDT track segments or the sagitta $s$. Both $\beta$ and $s$ are inversely proportional to the muon momentum  and depend on the magnetic field along the muon trajectory. The magnetic field inside the muon spectrometer has large non-uniformities which have to be taken into account in the momentum determination. It turns out that inside a spectrometer sector one can parametrize the dependence of transverse muon momentum $p_\mathrm{T}$ by a sum of three functions $S(\beta|s)$, $P(\phi)$, and $E(\eta)$ where $S$, $P$, adn $E$ are different for the use of $\beta$ or $S$ and $\phi$ and $\eta$ denote the polar angle and the pseudorapidity of the muon respectively:
\begin{eqnarray}
    p_\mathrm{T} = S(\beta|s)+P(\phi)+E(\eta),
    \nonumber\\
    S(\beta|s)=\alpha_1\cdot\left(\frac{1}{\beta|s}-a_0\right)
    \nonumber\\
    P(\phi)= p_0+p_1\cdot\phi+p_2\cdot\phi^2,
    \nonumber\\
    E(\eta)= e_0+e_1\cdot\eta+e_2\cdot\eta^2.
    \nonumber
\end{eqnarray}

\section{Latency estimates}

The overal latency of the MDT trigger has the following contributions:
\begin{itemize}
    \item   The time of flight of the muon from the interaction point to the
            MDT chambers. The time of flight is less than 65~$\mu$s.
    \item   The drift of the primary electrons in the gas of the tube. The      
            maximum drift time is 750~ns.
    \item   The time for the digitization of the MDT hits in the ADCs and TDCs.
    \item   The signal processing time in the on-chamber multiplexer.
    \item   The time for the transmission of the MDT data from the multiplexer
            to the off-chamber MDT trigger electronics, which takes 516~ns for
            the longest optical fibre of 100~m length.
    \item   The hit matching in the hit matching unit.
    \item   The time to transfer the matched hits to the MDT track 
            reconstruction unit.
    \item   The processing time in the MDT track reconstruction unit.
\end{itemize}

In order to estimate the time for the digitization of the MDT hits, the signal processing time in the on-chamber multiplexer, the hit matching and the MDT track reconstruction times demonstrators for the corresponding functional units were built. For the ron-chamber electronics we used the same ASD chips as in the present ATLAS detector, but replaced the TDC by a triggerless TDC implemented on an FPGA. Our on-chamber multiplexer was based on CERN's GLIB platform. A Xilinx Zynq evaluation board was used as demonstrator of the off-chamber trigger electronics. As the trigger chambers are almost 1~$\mu$s faster than the MDT chambers and the trigger algorithm is less complex than the MDT segment reconstruction algorithm, the muon pretrigger will be available already 1.5~$\mu$s after a $pp$ collision which is before the arrival of the latest MDT hit.

\begin{figure}[hbt]
\begin{center}
    \includegraphics[width=\linewidth]{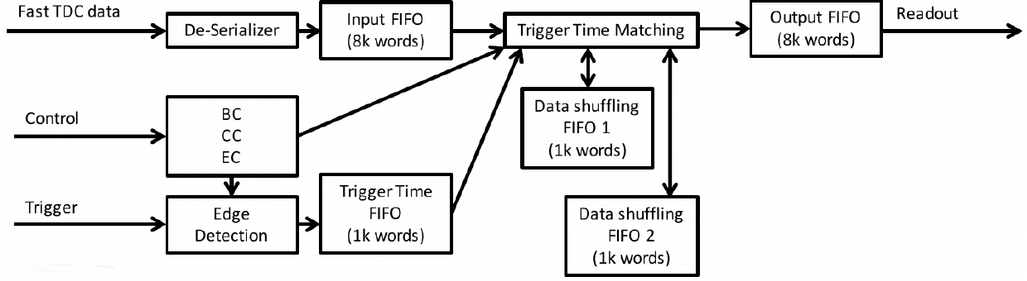}
    \caption{\label{sec:latency:fig01} Block diagram describing the 
    implementation of the demonstrator for the hit matching unit of the 
    off-chamber MDT trigger electronics.}
\end{center}
\end{figure}

Figure~\ref{sec:latency:fig01} shows a block diagram explaining the implementation of the hit matcher. The hit has three input: an input for the data from the on-chamber TDCs, an input for control data like bunch counter reset requests, and an input for the pretriggers. The MDT data are stored in an 8k words large input FIFO, the pretriggers in a 1k word large trigger time FIFO. The trigger time matching avoid pointers to the MDT hits and uses two data shuffling FIFOS. The trigger time matcher goes through all hits in the active data shuffling FIFO and checks if the MDT data are too old and expired. If they are not expired the MDT data are transferred to the other data shuffling FIFO. If the MDT hit can be matched to the current trigger, its data is sent to the output FIFO. On the trigger the same procedure is repeated this time looking at the data in the second data shuffling FIFO. 

The segment reconstruction algorithm was also implemented on the Zynq FPGA of the evaluation board. The pattern recognition part was implemented as a Xilinx IP core to handle 16 MDT hits in parallel which is the maximum number of muon hits in an MDT chamber. The $\sqrt{1+\bar{m}^2}$ term was replaced by its Taylor expansion up to $\bar{m}^2$ which can be computed based on summations and bit shifts. For the ordering of the $b_\pm$ values an implementation of a parallel bubble sort algorithm was used. The overal implementation consumes very little FPGA resources, indeed, it requires 3960 slice LUTs 1823 slice registers, 59 F7 muxes, and 40 DSPs. The MDT data selected by the pattern recognition unit were transferred over a 32-bit AXI bus at 320~MHz clock frequency to the embedded ARM Cortex A9 processor running at 800~MHz. The ARM processor receives one 32-bit header containing $\bar{m}$, one 32-bit word per hit which contains the sign of $b_\pm$, a unique tube identifier and the drift radius in multiples of 15~$\mu$m. The segment fit is performed on the ARM processor on an interrupt sent by the pattern recognition unit. Finally the segment parameters are used for the momentum determination.

\begin{table}[hbt]
\begin{center}
\caption{\label{sec:latency:tab01}Summary of the contributions to the MDT trigger latency based on the MDT trigger demonstrators.}
\begin{tabular}{|l|c|c|}
\hline
Step    & Latency   & Comment \\
\hline
Time of flight  & 65~ns & \\
Maximum drift time  & 750~ns & \\
Digitization and multiplexing   & 561~ns & \\
Data transfer to the trigger & & \\
electronics & 516~ns & \\
Hit matching    & 440~ns    & \\
Transfer of matched hits to the & & \\
track reconstruction unit & 250~ns & Assuming GBT link \\
Pattern recognition & 204~ns & 250~MHz FPGA \\
 & & clock frequency \\
Transfer of hit pattern to & & \\
ARM processor    &  60~ns & 800~MHz ARM \\
Segment fit &   500~ns  & clock frequency \\
Transfer of segment parameters & & \\
to Zynq FPGA on mother board    & 250~ns & Assuming GBT link \\
Momentum determination & 80~ns & \\
\hline
{\bf Total} & {\bf 3626~ns} & \\
\hline
\end{tabular}
\end{center}
\end{table}

Table~\ref{sec:latency:tab01} summarizes the contributions to the overall latency of the MDT trigger. The signal processing time in the on-chamber electronics and the MDT hit matching unit were determined with test pulses. The latency of the track reconstruction unit was determined with simulated data. It turns out that the muon momentum based on the MDT data can be provided after 3.6~$\mu$s after a $pp$ collision which is well within the ATLAS L0 trigger latency of 10~$\mu$s.

\section{Summary}
A highly selective muon trigger with a sharp trigger turn-on curve is mandatory for the operation of the ATLAS detector at the HL-LHC. This requires the use of the precision MDT chamber data already in the first trigger level. We have described and demonstrated a concept for the MDT trigger which will make it possible to provide an MDT based first-level muon trigger well within the latency of the first-level trigger of 10~$\mu$s.


%





\ifCLASSOPTIONcaptionsoff
  \newpage
\fi

\end{document}